\documentclass[10pt]{article}
\usepackage{bbm}
\usepackage[final]{graphics}
\usepackage{amsfonts,amsbsy}

\font\tenimbf=cmmib10 at 10pt
\font\sevenimbf=cmmib10 at 7pt
\font\fiveimbf=cmmib10 at 5pt
\newfam\imbf
\textfont\imbf=\tenimbf
\scriptfont\imbf=\sevenimbf
\scriptscriptfont\imbf=\fiveimbf

\def\empile#1\over#2{\mathrel{\mathop{\kern 0pt#1}\limits_{#2}}}

\newcommand{\slv}{\raise.15ex\hbox{$/$}\kern-.53em\hbox{$v$}}
\newcommand{\slF}{\raise.15ex\hbox{$/$}\kern-.53em\hbox{$F$}}
\newcommand{\slL}{\raise.15ex\hbox{$/$}\kern-.53em\hbox{$L$}}
\newcommand{\slP}{\raise.15ex\hbox{$/$}\kern-.53em\hbox{$P$}}
\newcommand{\slp}{\raise.15ex\hbox{$/$}\kern-.53em\hbox{$p$}}
\newcommand{\slq}{\raise.15ex\hbox{$/$}\kern-.53em\hbox{$q$}}
\newcommand{\slR}{\raise.15ex\hbox{$/$}\kern-.53em\hbox{$R$}}
\newcommand{\slQ}{\raise.15ex\hbox{$/$}\kern-.53em\hbox{$Q$}}
\newcommand{\slK}{\raise.15ex\hbox{$/$}\kern-.53em\hbox{$K$}}
\newcommand{\slk}{\raise.15ex\hbox{$/$}\kern-.53em\hbox{$k$}}
\newcommand{\slD}{\raise.15ex\hbox{$/$}\kern-.53em\hbox{$D$}}
\newcommand{\slA}{\raise.15ex\hbox{$/$}\kern-.53em\hbox{$A$}}
\newcommand{\slG}{\raise.15ex\hbox{$/$}\kern-.53em\hbox{$G$}}
\newcommand{\slSigma}{\raise.15ex\hbox{$/$}\kern-.53em\hbox{$\Sigma$}}
\newcommand{\slpartial}{\raise.15ex\hbox{$/$}\kern-.53em\hbox{$\partial$}}
\newcommand{\slcalP}{\raise.15ex\hbox{$/$}\kern-.63em\hbox{$\cal P$}}

\def\p{{\boldsymbol p}}
\def\q{{\boldsymbol q}}

\def\k{{\boldsymbol k}}

\def\x{{\boldsymbol x}}

\begin{document}
\thispagestyle{empty}

\title{\bf Remarks on transient photon production in heavy ion collisions}
\author{Eduardo Fraga$^{(1)}$, Fran\c cois Gelis$^{(2)}$, Dominique Schiff$^{(3)}$}
\maketitle
\begin{center}
\begin{enumerate}
\item Intituto de F\'\i sica\\
Universidade Federal do Rio de Janeiro\\
C.P. 68528, Rio de Janeiro, RJ 21941-972, Brazil
\item Service de Physique Th\'eorique\\
CEA/DSM/Saclay, Orme des merisiers\\
91191 Gif-sur-Yvette cedex, France
\item Laboratoire de Physique Th\'eorique\\
B\^at. 210, Universit\'e Paris XI\\
91405 Orsay cedex, France
\end{enumerate}
\end{center}

%%%%%%%%%%%%%%%%%%%%%%%%%%%%%%%%%%%%%%%%%%%%%%%%

\begin{abstract}
  In this note, we discuss the derivation of a formula that has been
  used in the literature in order to compute the number of photons
  emitted by a hot or dense system during a finite time. Our
  derivation is based on a variation of the standard operator-based
  $S$-matrix approach. The shortcomings of this formula are then
  emphasized, which leads to a negative conclusion concerning the
  possibility of using it to predict transient effects for the photon
  rate.
\end{abstract}

%%%%%%%%%%%%%%%%%%%%%%%%%%%%%%%%%%%%%%%%%%%%%%%%

\vspace{1cm}

\section{Introduction}
Electromagnetic radiation (photons and lepton pairs) has long been
thought to be a good probe of the early stages of heavy ion
collisions.  Indeed, the production rates of these particles
increase very rapidly with temperature and therefore are dominated by
the early times. In addition, photons and leptons are interacting very
weakly with quarks and gluons, which means that final state
interactions can in general be neglected.

The photon and dilepton production rates have been evaluated
completely at leading order in the strong coupling $\alpha_s$ for a
quark-gluon plasma in {\sl local thermal equilibrium}.  These rates
have been evaluated at $1$-loop\footnote{The loop counting refer to
diagrams of the effective theory one obtains after resumming the hard
thermal loops \cite{BraatP1,BraatP2}.}  in
\cite{KapusLS1,BaierNNR1,AltheR1}, at $2$-loop in
\cite{AurenGKP1,AurenGKP2,AurenGKZ1,AurenGZ2}, and the
Landau-Pomeranchuk-Migdal \cite{LandaP1,LandaP2,Migda1} corrections
have been resummed in
\cite{ArnolMY1,ArnolMY2,ArnolMY3,Sastr1,Sastr2,AurenGZ4,AurenGMZ1}. Up-to-date
reviews of the situation regarding photon and dilepton production in
equilibrium can be found in \cite{Gelis13,GaleH1,KraemR1,Arleoa1}.

The situation is in a far less advanced state when the system is not
in local thermal equilibrium. Some attempts have been made in order to
take into account in the calculation the effect of a departure from
chemical equilibrium. This is done by introducing {\sl fugacities} in
the quark and gluon distributions. The fugacity dependent rates have
been evaluated at $1$-loop in \cite{BaierDRS1}, at $2$-loop in
\cite{DuttaSMKC1}, and generalized to the infinite series of diagrams
that contribute to the LPM effect in \cite{Arleoa1}. These
calculations are based on a minimal extension of the equilibrium
formalism, that merely consists in replacing the equilibrium
statistical distributions by non-equilibrium distributions. As a
consequence, they may suffer from the so-called pinch singularities.
It can however be shown that the corrections due to the ``pinch
terms'' are negligible if the process under study is fast compared to
the relaxation time \cite{GelisSS1}.

In recent years, a new ``real-time'' approach has been proposed in
order to compute out-of-equilibrium -- namely time-dependent --
effects for photon production in a dense equilibrated quark-gluon
system, which originate in its finite life-time \cite{WangB1,WangBN1}.
The result was that there are important transient effects that make
the yield much larger than what would have been expected by simply
multiplying the equilibrium rates by the corresponding amount of time.

This unexpectedly large photon yield, combined with the fact that the
energy spectrum is not integrable\footnote{This appears to have
  changed in more recent iterations \cite{BoyanV1,BoyanV2} of this
  work, thanks to a subtraction of the terms responsible for this
  behavior.}, was the starting point of many discussions regarding the
validity of this approach \cite{Moore1,Serre2,Arleoa1}. In this note,
we critically discuss the assumptions underlying the formulas used in
\cite{WangB1,WangBN1,BoyanV1}. To that effect, we propose a derivation
of the formula giving the photon yield which is based on the standard
canonical formalism. In particular, we show that it can be obtained
from an ``almost standard'' $S$-matrix approach, the only departure
from the standard being that we need to turn on and off the
electromagnetic interactions at some finite times $t_i$ and $t_f$.

Our paper is organized as follows. Section $2$ is devoted to a
derivation of the photon yield up to the order $e^2$ of an expansion
in the electromagnetic coupling constant. In section $3$, we come
back to the assumptions made in the derivation of the previous
section, and discuss their relevance and validity. Section $4$ is
devoted to concluding remarks.

\section{Photon production}
\subsection{General framework}
Let us now consider a system of quarks and gluons, and denote by
$H_{_{QCD}}$ its Hamiltonian (containing all the QCD interactions). We
couple the quarks to the electromagnetic field in order to study
photon emission by this system, and denote $H_{\rm e.m.}$ the
Hamiltonian of the electromagnetic field, and $H_{q\gamma}$ the term
of the Hamiltonian that couples the quarks to the photons. The
complete Hamiltonian is therefore:
\begin{equation}
H=H_{_{QCD}}+H_{\rm e.m.}+H_{q\gamma}\; .
\end{equation}
At the level of the Lagrangian, this reads:
\begin{eqnarray}
&&{\cal L}={\cal L}_{_{QCD}}+{\cal L}_{\rm e.m.}+{\cal L}_{q\gamma}\; ,
\nonumber\\
&&{\cal L}_{_{QCD}}\equiv -\frac{1}{4}G_{\mu\nu}^a G^{\mu\nu}_a
+\overline{\psi}(i\slpartial_x-g\slG(x)-m)\psi(x)\; ,
\nonumber\\
&&{\cal L}_{\rm e.m.}\equiv -\frac{1}{4}F_{\mu\nu}F^{\mu\nu}\; ,
\nonumber\\
&&{\cal L}_{q\gamma}\equiv -e\overline{\psi}(x)\slA(x)\psi(x)\; ,
\end{eqnarray}
where $G_\mu$ and $G_{\mu\nu}^a$ are respectively the gluon field and
field strength, $A_\mu$ and $F_{\mu\nu}$ the photon field and field
strength, and $\psi$ the quark field (only one flavor is considered
here). $g$ is the strong coupling constant, and $e$ is the quark
electric charge. We denote collectively by ${\cal L}_{\rm int}$ the
sum of all the interaction terms.

The number of photons measured in the system at some late time is
given by the following formula:
\begin{eqnarray}
2\omega \frac{dN}{d^3\x d^3\k}=\frac{1}{V}
\sum_{{\rm pol.\ }\lambda}\;\frac{1}{Z}
{\rm Tr}\Big(\rho(t_i)a^{(\lambda)\dagger}_{\rm out}(\k)
a^{(\lambda)}_{\rm out}(\k)\Big)\; .
\label{eq:N1}
\end{eqnarray}
$V$ is the volume of the system and $Z\equiv{\rm Tr}(\rho(t_i))$ the
partition function. The sum runs over the physical polarization states
of the photon.  $\rho(t_i)$ is the density operator that defines the
initial statistical ensemble. The ``${\rm in}$'' states and operators
of the interaction picture are free, and are defined to coincide with
those of the Heisenberg picture at the initial time $t=t_i$. The
``${\rm out}$'' states and operators are those used in order to
perform the measurement. In principle, the measurement should take
place after the photons have stopped interacting, i.e. one should
count the photons at a time $t\to +\infty$ so that they are
asymptotically free photons.  Here, for the sake of the argument, we
are going to define the ``${\rm out}$'' states and fields at some
finite time $t_f$. This means that we assume that electromagnetic
interactions have been turned off before the time $t_f$, for this
measurement to be meaningful.  These ``{\rm out}'' states and fields
are related to the ``${\rm in}$'' states and fields by means of the
``$S$-matrix'':
\begin{eqnarray}
&&\big|\alpha_{\rm out}\big>=S^\dagger\big|\alpha_{\rm in}\big>\; ,\nonumber\\
&&a^{(\lambda)}_{\rm out}(\k)=S^\dagger a^{(\lambda)}_{\rm in}(\k) S\; ,
\nonumber\\
&&a^{(\lambda)\dagger}_{\rm out}(\k)=S^\dagger a^{(\lambda)\dagger}_{\rm in}(\k) S\; ,
\label{eq:out}
\end{eqnarray}
with an $S$ matrix given in terms of the interaction as\footnote{Note
that this is the $S$-matrix for a system in which the interactions are
switched on at the time $t_i$ and switched off at the the time
$t_f$.}:
\begin{eqnarray}
S&=&U(t_f,t_i)\equiv{\cal P}\exp i\int_{t_i}^{t_f}d^4x 
{\cal L}_{\rm int}(\phi_{\rm in}(x))\; ,
\nonumber\\
S^\dagger&=&U(t_i,t_f)\equiv{\cal P}\exp i\int_{t_f}^{t_i}d^4x 
{\cal L}_{\rm int}(\phi_{\rm in}(x))\; .
\end{eqnarray}
${\cal P}$ denotes the path-ordering. Expressing the ``out'' creation
and annihilation operators in terms of their ``in'' counterparts, we
have:
\begin{eqnarray}
2\omega\frac{dN}{d^3\x d^3\k}=\frac{1}{V}\sum_{{\rm pol.\ }\lambda}
\;\frac{1}{Z}{\rm Tr}\,\left(
\rho(t_i) S^\dagger a^{(\lambda)\dagger}_{\rm in}(\k) a^{(\lambda)}_{\rm in}(\k) S
\right)\; .
\label{eq:photon-number-1}
\end{eqnarray}
The next step required in order to bring this expression to an easily
calculable form is to rewrite the creation and annihilation operators
in terms of the corresponding fields\footnote{The interaction picture
field $A^\mu_{{\rm in}}(x)$ being a free field, these relations indeed
give creation and annihilation operators that are time
independent. One can therefore choose arbitrarily the value of $x_0$
when evaluating Eq.~(\ref{eq:rel-a-A}).}:
\begin{eqnarray}
&&a^{(\lambda)\dagger}_{\rm in}(\k)=-i\varepsilon_\mu^{(\lambda)*}(\k)
\int d^3\x e^{-ik\cdot x} 
\stackrel{\leftrightarrow}{\partial}_{x_0} 
A^\mu_{\rm in}(x)\; ,\nonumber\\
&&a^{(\lambda)}_{\rm in}(\k)=i\varepsilon_\mu^{(\lambda)}(\k)
\int d^3\x e^{ik\cdot x} 
\stackrel{\leftrightarrow}{\partial}_{x_0} 
A^\mu_{\rm in}(x)\; ,
\label{eq:rel-a-A}
\end{eqnarray}
where $\varepsilon_\mu^{(\lambda)}(\k)$ is the polarization vector for
a photon of momentum $\k$ and polarization $\lambda$.  The symbol
$\stackrel{\leftrightarrow}{\partial}_{x_0}$ is defined as follows:
\begin{equation}
A(x_0,\x)\stackrel{\leftrightarrow}{\partial}_{x_0} B(x_0,\x)
\equiv
A(x_0,\x)\left(\partial_{x_0} B(x_0,\x)\right)-
\left(\partial_{x_0} A(x_0,\x)\right)B(x_0,\x)\; .
\end{equation}
Using the property $U(t_1,t_2)U(t_2,t_3)=U(t_1,t_3)$ and the previous
relations, we can rewrite Eq.~(\ref{eq:photon-number-1}) as follows:
\begin{eqnarray}
&&2\omega\frac{dN}{d^3\x d^3\k}=\frac{1}{V}\sum_{{\rm pol.\ }\lambda}
\varepsilon^{(\lambda)*}_\mu(\k)\varepsilon^{(\lambda)}_\nu(\k)
e^{i\omega\cdot(y_0-x_0)} 
\stackrel{\leftrightarrow}{\partial}_{x_0}
\stackrel{\leftrightarrow}{\partial}_{y_0}
\nonumber\\
&&\qquad\times\frac{1}{Z}
{\rm Tr}\,\left[
\rho(t_i)U(t_i,t_f)
A^\mu_{\rm in}(x_0,\k)A^\nu_{\rm in}(y_0,-\k)
U(t_f,t_i)
\right]\, ,\nonumber\\
&&
\label{eq:photon-number-2}
\end{eqnarray}
where we have performed a Fourier transform on the spatial dependence of
the photon field:
\begin{equation}
A^\mu_{\rm in}(x_0,\k)\equiv \int d^3\x \,
e^{i\k\cdot\x} A^\mu_{\rm in}(x_0,\x)\; .
\end{equation}
In Eq.~(\ref{eq:photon-number-2}), one still has the freedom to chose
at will the times $x_0$ and $y_0$. This freedom can be exploited in
order to write:
\begin{eqnarray}
&&2\omega\frac{dN}{d^3\x d^3\k}=\frac{1}{V}
\sum_{{\rm pol.\ }\lambda}
\varepsilon^{(\lambda)*}_\mu(\k)\varepsilon^{(\lambda)}_\nu(\k)
({\partial}_{x_0}+i\omega)
({\partial}_{y_0}-i\omega)
\nonumber\\
&&\qquad\times\frac{1}{Z}
{\rm Tr}\,\left[
\rho(t_i){\cal P}\left(
A^{\mu,(-)}_{\rm in}(x_0,\k)A^{\nu,(+)}_{\rm in}(y_0,-\k)
e^{i\int_{\cal C}{\cal L}_{\rm int}}
\right)
\right]_{x_0=y_0=t_f}\, .\nonumber\\
&&
\label{eq:photon-number-3}
\end{eqnarray}
In this formula, the time path ${\cal C}$ goes from $t_i$ to $t_f$
along the real axis, and then back to $t_i$. Although it could be
included in the definition of ${\cal C}$, the portion of the contour
that goes from $t_f$ to $+\infty$ and back to $t_f$ would not
contribute to the expression (because
$U(t_f,+\infty)U(+\infty,t_f)=1$).  In Eq.~(\ref{eq:photon-number-3}),
the time derivatives act only on the right (and should be taken before
the times $x_0$ and $y_0$ are set equal to $t_f$). The subscript $(-)$
(resp. $(+)$) indicates that the corresponding field lives on the
lower (resp. upper) branch of the time-path; they are necessary in
order to ensure that the two fields remain correctly ordered when
acted upon by the path ordering operator.

\subsection{Order $e^0$}
At this point, one can perform an expansion in the electromagnetic
coupling constant, while conserving strong interactions to all
orders. This is motivated by the very different magnitude of the
electromagnetic and the strong coupling constants.  The term of
order 0 in the electric charge $e$ is obtained by replacing the full
interaction Lagrangian ${\cal L}_{\rm int}$ by the QCD interactions
only (which we denote ${\cal L}_{\rm int}^{^{QCD}}$), which leads to:
\begin{eqnarray}
&&2\omega\left.\frac{dN}{d^3\x d^3\k}\right|_{e^0}=\frac{1}{V}
\sum_{{\rm pol.\ }\lambda}
\varepsilon^{(\lambda)*}_\mu(\k)\varepsilon^{(\lambda)}_\nu(\k)
({\partial}_{x_0}+i\omega)
({\partial}_{y_0}-i\omega)\nonumber\\
&&\qquad\qquad\qquad\qquad\times
\left<A^{\mu,(-)}_{\rm in}(x_0,\k)A^{\nu,(+)}_{\rm in}(y_0,-\k)\right>_{x_0=y_0=t_f}
\, ,
\label{eq:tmp-order-0}
\end{eqnarray}
where we denote:
\begin{equation}
\left<{\cal A}\right>\equiv
\frac{
{\rm Tr}\,\left[
\rho(t_i){\cal P}
{\cal A}\,
e^{i\int_{\cal C}{\cal L}^{^{QCD}}_{\rm int}}
\right]}
{{\rm Tr}\,\left[
\rho(t_i)
\right]}
\end{equation}
the ensemble average of an operator ${\cal A}$ at order zero in the
electromagnetic coupling (but to all orders in the strong coupling
constant).  Since this object is evaluated at order 0 in the
electromagnetic coupling constant, the correlator that appears in
eq.~(\ref{eq:tmp-order-0}) is nothing but the {\sl free} path-ordered
photon propagator (the exponential containing the strong interactions
drops out because the photons do not couple to quarks at this
order\footnote{Quarks and gluons can only enter in disconnected
vacuum-vacuum diagrams that are zero when the time integrations are
carried out on both branches of the closed time path ${\cal
C}$.}). Therefore, we have in the Feynman gauge:
\begin{eqnarray}
&&
\left<
A^{\mu,(-)}_{\rm in}(x_0,\k)A^{\nu,(+)}_{\rm in}(y_0,-\k)
\right>
=\nonumber\\
&&\qquad=-g^{\mu\nu}\frac{V}{2\omega}\left[
(1+n_\gamma(\omega))e^{-i\omega(x_0-y_0)}
+n_\gamma(\omega)e^{i\omega(x_0-y_0)}
\right]\; ,
\label{eq:2phot-pm}
\end{eqnarray}
where we have used explicitly the fact that $x_0$ is always posterior
to $y_0$ on the time-path. We denote by $n_\gamma(\omega)$ the
statistical distribution of photons present in the initial ensemble
described by $\rho(t_i)$. The prefactor $V$ comes from translation
invariance: the correlator is proportional to a momentum conservation
delta function $(2\pi)^3\delta(\k-\k)=V$. Applying the operator
$({\partial}_{x_0}+i\omega) ({\partial}_{y_0}-i\omega)$ on the
previous correlation function, we obtain trivially:
\begin{eqnarray}
\left.\frac{dN}{d^3\x d^3\k}\right|_{e^0}
=-\left[\sum_{{\rm pol.\ }\lambda} 
\varepsilon^{(\lambda)*}_\mu(\k)\varepsilon^{(\lambda)}{}^\mu(\k)
\right]\,n_\gamma(\omega)=2n_\gamma(\omega)\; .
\end{eqnarray}
Naturally, this result was expected: at order ${\cal O}(e^0)$, the
number of photons in the system is given by the initial photon
distribution function multiplied by the number of physical degrees of
polarization. This is zero if we assume that the system does not
contain any photons initially.

\subsection{Order $e^2$}
Expanding $\exp -ie\int \overline{\psi}\slA\psi$ to order $e^1$
produces a 3-photon correlator, which vanishes by Furry's theorem.
Therefore, the next non zero contribution can occur only at the order
$e^2$. Expanding the exponential containing the electromagnetic
interactions up to this order, we obtain\footnote{Implicitly, this is
  the correct $e^2$ expansion only if the initial density matrix does
  not depend on the coupling constant $e$, i.e. if there are no
  electromagnetic interactions in the system at the initial time.}:
\begin{eqnarray}
&&2\omega
\left.\frac{dN}{d^3\x d^3\k}\right|_{e^2}
=-\frac{e^2}{2V}
\sum_{{\rm pol.\ }\lambda}
\varepsilon^{(\lambda)*}_\mu(\k)\varepsilon^{(\lambda)}_\nu(\k)
({\partial}_{x_0}+i\omega)
({\partial}_{y_0}-i\omega)
\nonumber\\
&&\qquad\qquad\qquad\qquad\qquad\times
\left<A^{\mu,(-)}_{\rm in}(x_0,\k)A^{\nu,(+)}_{\rm in}(y_0,-\k)
L_2\right>\; ,
\end{eqnarray}
with the shorthand:
\begin{equation}
L_2\equiv
\int_{\cal C}du_0\,dv_0\int\frac{d^3\p}{(2\pi)^3}\frac{d^3\q}{(2\pi)^3}
A^\rho_{\rm in}(u_0,\p)A^\sigma_{\rm in}(v_0,\q)
J_{{\rm in},\rho}(u_0,-\p)J_{{\rm in}, \sigma}(v_0,-\q)\; .
\end{equation}
$J_{{\rm in},\rho}=\overline{\psi}_{\rm in}\gamma_\rho \psi_{\rm in}$
is the electromagnetic vector current.  The correlator that appears in
this formula can now be expanded using Wick's theorem. Naturally, the
fermionic currents can only be paired among themselves. One of the
pairings corresponds to a term that contains a disconnected
vacuum-vacuum factor: such a vacuum-vacuum diagram is zero when
evaluated on the closed time path ${\cal C}$. We are left with:
\begin{eqnarray}
&&\big<
A^{\mu,(-)}_{\rm in}(x_0,\k)A^{\nu,(+)}_{\rm in}(y_0,-\k) L_2
\big>
\nonumber\\
&&=\int_{\cal C}du_0\,dv_0\int\frac{d^3\p}{(2\pi)^3}\frac{d^3\q}{(2\pi)^3}
\big<J_{{\rm in},\rho}(u_0,-\p)J_{{\rm in}, \sigma}(v_0,-\q)\big>
\nonumber\\
&&\qquad\qquad\times
\Big\{\big<A^{\mu,(-)}_{\rm in}(x_0,\k)A^\rho_{\rm in}(u_0,\p)\big>
\big<A^{\nu,(+)}_{\rm in}(y_0,-\k)A^\sigma_{\rm in}(v_0,\q)\big>
\nonumber\\
&&\qquad\qquad\;\;
+\big<A^{\mu,(-)}_{\rm in}(x_0,\k)A^\sigma_{\rm in}(v_0,\q)\big>
\big<A^{\nu,(+)}_{\rm in}(y_0,-\k)A^\rho_{\rm in}(u_0,\p)\big>
\Big\}\; .
\label{eq:wick}
\end{eqnarray}
The 2-photon correlator is:
\begin{eqnarray}
\left<
A^\mu_{\rm in}(x_0,\k)A^\nu_{\rm in}(y_0,\k^\prime)
\right>=
-(2\pi)^3\delta(\k+\k^\prime)g^{\mu\nu} G(x_0,y_0;\k)\; ,
\end{eqnarray}
where we denote
\begin{eqnarray}
&&G(x_0,y_0;\k)\equiv\frac{1}{2\omega}\big[
(\theta_c(x_0\!-\!y_0)\!+\!n_\gamma(\omega))e^{-i\omega(x_0-y_0)}
\nonumber\\
&&\qquad\qquad\qquad\qquad
+
(\theta_c(y_0\!-\!x_0)\!+\!n_\gamma(\omega))e^{-i\omega(y_0-x_0)}
\big]
\end{eqnarray}
with $\theta_c(x_0-y_0)$ the generalization of the step function to
the time path ${\cal C}$. This formula is a generalization of
eq.~(\ref{eq:2phot-pm}) to the case where the times $x_0$ and $y_0$
can lie anywhere on the time path. The current-current correlator can
be written in terms of the photon polarization tensor as follows:
\begin{eqnarray}
\big<J_{{\rm in},\rho}(u_0,-\p)J_{{\rm in}, \sigma}(v_0,-\q)\big>
=(2\pi)^3\delta(\p+\q)\Pi_{\rho\sigma}(u_0,v_0;-\p)\; .
\end{eqnarray}
Acting on eq.~(\ref{eq:wick}) with the operator
$({\partial}_{x_0}+i\omega) ({\partial}_{y_0}-i\omega)$, we obtain
easily:
\begin{eqnarray}
&&2\omega\left.\frac{dN}{d^3\x d^3\k}\right|_{e^2}
=-\frac{e^2}{2}
\sum_{{\rm pol.\ }\lambda}
\varepsilon^{(\lambda)*}_\mu(\k)\varepsilon^{(\lambda)}_\nu(\k)
\int_{{\cal C}}du_0\,dv_0\,e^{i\omega(v_0-u_0)}\nonumber\\
&&\qquad\qquad\times
[n_\gamma(\omega)+\theta_c(u_0-t_f^-)]
[n_\gamma(\omega)+\theta_c(t_f^+-v_0)]
\Pi^{\mu\nu}(u_0,v_0;\k)\; ,
\end{eqnarray}
where the superscript $+$ or $-$ on the time $t_f$ indicates on which
branch of the contour ${\cal C}$ the corresponding time must be
considered.

At this point, we can break down the contour integrations in domains
where the $\theta$ functions have a constant value. Using the property:
\begin{equation}
\Pi^{\mu\nu}_{++}(u_0,v_0;\k)+\Pi^{\mu\nu}_{--}(u_0,v_0;\k)
=
\Pi^{\mu\nu}_{+-}(u_0,v_0;\k)+\Pi^{\mu\nu}_{-+}(u_0,v_0;\k)
\end{equation}
which is valid even out-of-equilibrium, we obtain:
\begin{eqnarray}
&&2\omega\left.\frac{dN}{d^3\x d^3\k}\right|_{e^2}
=-\frac{e^2}{2}
\sum_{{\rm pol.\ }\lambda}
\varepsilon^{(\lambda)*}_\mu(\k)\varepsilon^{(\lambda)}_\nu(\k)
\int\limits_{t_i}^{t_f}du_0\,dv_0\,e^{i\omega(v_0-u_0)}\nonumber\\
&&\qquad\qquad\qquad\times
[n_\gamma(\omega)\Pi^{\mu\nu}_{+-}(u_0,v_0;\k)
-(1+n_\gamma(\omega))\Pi^{\mu\nu}_{-+}(u_0,v_0;\k)
]\; .
\end{eqnarray}
Performing the sum over the photon polarizations, and combining the
order $e^0$ and the order $e^2$ results, we have the following photon
distribution at time $t_f$:
\begin{eqnarray}
&&2\omega\frac{dN}{d^3\x d^3\k}
=2n_\gamma(\omega)
+\frac{e^2}{2}\int\limits_{t_i}^{t_f}du_0dv_0\,e^{i\omega(v_0-u_0)}\nonumber\\
&&\qquad\qquad\qquad\times
[n_\gamma(\omega)\Pi^{\mu}_\mu{}_{+-}(u_0,v_0;\k)
-(1+n_\gamma(\omega))\Pi^{\mu}_\mu{}_{-+}(u_0,v_0;\k)
]
\nonumber\\
&&
\qquad\qquad\qquad\qquad\quad\;
+{\cal O}(e^4)\; .
\label{eq:ngamma-final}
\end{eqnarray}
This formula can be applied to several particular cases, that we
discuss in the following sections.

\section{Photons, quarks and gluons in thermal equilibrium}
A first situation to consider is that of an initial ensemble which
corresponds to quarks, gluons, {\sl and photons} in thermal
equilibrium at a given temperature $T$. This means that the initial
distributions of these particles are given by the Fermi-Dirac and
Bose-Einstein functions at temperature $T$. Moreover, the initial
density operator is $\rho(t_i)=\exp(-H/T)$ where $H$ is the full
Hamiltonian of the system, {\sl including all the interactions (strong
  as well as electromagnetic).} In this case\footnote{When
  $\rho(t_i)=\exp(-H/T)$, there is an explicit dependence of the
  initial density matrix on the coupling constant $e$. It is known
  that this dependence is taken care of without changing any formula,
  simply by appending a vertical branch to the time path
  \cite{Gelis1}, going from $t_i$ to $t_i-i/T$. Moreover, it has been
  shown that the contribution of this vertical branch can be taken
  into account if one enforces the KMS condition at each intermediate
  step of the calculation \cite{Gelis1,BellaM1,Gelis8}.}, the photon
polarization tensor obeys the KMS identity, which reads:
\begin{equation}
n_\gamma(\omega)\Pi^{\mu\nu}_{+-}(u_0,v_0;\k)=
(1+n_\gamma(\omega))\Pi^{\mu\nu}_{-+}(u_0,v_0;\k)\; .
\end{equation}
This relation implies that the $e^2$ term in
eq.~(\ref{eq:ngamma-final}) vanishes. One could in fact check that all
the higher order terms (in $e^2$) in the photon distribution would
also vanish thanks to the KMS identity. Naturally, this result is
expected: if the initial ensemble is an equilibrium ensemble, the
populations of particles do not change over time. It is also natural
that in the calculation this property appears as a consequence of KMS:
indeed, KMS is the manifestation of thermal equilibrium at the level
of the Green's functions. In practice, the situation considered in
this section is only of academic interest, because in applications
such as heavy ion collisions the photons are not in equilibrium with
the strongly interacting particles.

\section{Photon-free initial state}
More interesting is the situation of a system which does not contain
any real photons initially, i.e. for which $n_\gamma(\omega)=0$. In
this case, the photon population at time $t_f$ is given by:
\begin{equation}
2\omega\frac{dN}{d^3\x d^3\k}
=-\frac{e^2}{2}
\int\limits_{t_i}^{t_f}du_0\,dv_0\,e^{i\omega(v_0-u_0)}
\Pi^{\mu}_\mu{}_{-+}(u_0,v_0;\k)
\; .
\end{equation}
 If the initial density matrix describing the distribution of quarks
and gluons is such that $\Pi^{\mu}_\mu{}_{-+}(u_0,v_0;\k)$ is
invariant under time translation, we can introduce the Fourier
transform of the photon polarization tensor:
\begin{equation}
\Pi^{\mu}_\mu{}_{-+}(u_0,v_0;\k)
\equiv
\int\limits_{-\infty}^{+\infty} \frac{dE}{2\pi}e^{iE(u_0-v_0)}
\Pi^{\mu}_\mu{}_{-+}(E,\k)\; ,
\end{equation}
and we can write:
\begin{eqnarray}
2\omega\frac{dN}{d^3\x d^3\k}
=-e^2\int\limits_{-\infty}^{+\infty}
\frac{dE}{2\pi}
\frac{1-\cos((E-\omega)(t_f-t_i))}{(E-\omega)^2}
\Pi^{\mu}_\mu{}_{-+}(E,\k)\; .
\end{eqnarray}
If one takes the derivative with respect to $t_f$, one obtains the
number of photons produced per unit time and per unit phase-space at
the time $t_f$:
\begin{eqnarray}
2\omega\frac{dN}{dtd^3\x d^3\k}
=-e^2\int\limits_{-\infty}^{+\infty}
\frac{dE}{2\pi}
\frac{\sin((E-\omega)(t_f-t_i))}{E-\omega}
\Pi^{\mu}_\mu{}_{-+}(E,\k)\; .
\label{eq:photon-rate}
\end{eqnarray}
This formula is equivalent to the formula obtained by Boyanovsky et
al. However, we have derived it here within the framework of an
$S$-matrix formulation. This was not the standard $S$-matrix approach
though, as some extra hypothesis and some extensions have been used.
They are discussed in the next section.

\section{Discussion}
A first consequence of eq.~(\ref{eq:photon-rate}) is that it gives
back the usual formula for the photon production rate at equilibrium
if one takes to infinity the time $t_f$ at which the measurement is
performed (which amounts to turn off adiabatically the electromagnetic
interactions only at asymptotic times). Indeed, one has the
following property:
\begin{equation}
\lim_{t_f\to+\infty} \frac{\sin((E-\omega)(t_f-t_i))}{E-\omega}
=\pi\delta(E-\omega)\; ,
\end{equation}
which implies:
\begin{eqnarray}
2\omega\frac{dN}{dtd^3\x d^3\k}
=-\frac{e^2}{2}
\Pi^{\mu}_\mu{}_{-+}(\omega,\k)\; ,
\end{eqnarray}
i.e. we recover in this limit the usual relation between the photon
production rate and the on-shell photon polarization tensor. In
particular, there is no contribution to the rate at zeroth order in
$\alpha_s$.

At this point, the main question is whether the finite $t_f$
generalization of this formula makes sense as a photon production
rate, as invoked in \cite{BoyanV1}. In order to discuss this
possibility, it is useful to recall and discuss here the hypothesis
that have been used in order to derive eq.~(\ref{eq:photon-rate}).
\begin{itemize}
\item One may be tempted to interpret the number operator
  $a^\dagger_{\rm out}a_{\rm out}$ we have defined in the third of
  eqs.~(\ref{eq:out}) as the number operator at the time $t_f$ for
  photons still interacting with the system, but there is no warranty
  that this definition of the number of photons agrees with the number
  of photons as measured in a detector, precisely because they are not
  asymptotically free states.
  
  The only possibility to argue safely that this operator indeed
  counts observable photons is to assume that the system does not
  undergo electromagnetic interactions after the time $t_f$. Another
  way to state this is to say that the object $S$ which appears in
  eq.~(\ref{eq:out}) is the standard $S$-matrix connecting free ``out''
  and ``in'' states only if there are no interactions after $t_f$.
  
  In any case, it is clearly unphysical to keep $t_f$ finite: either
  we are trying to measure non asymptotically free photons, or we have
  to turn off the interactions at a finite time $t_f$.
  
\item A similar problem arises at the initial time. The derivation we
  have used for eq.~(\ref{eq:photon-rate}) assumed that there is no
  dependence on $e$ in the initial density operator $\rho(t_i)$, which
  is possible only if there are no electromagnetic interactions in the
  initial state.  In particular, imposing $n_\gamma=0$ in the initial
  state is also forbidding electromagnetic interactions before $t_i$
  (because a system of interacting quarks and gluons that undergo
  electromagnetic interactions will necessarily contain photons as
  well).
  
\item It is also known that eq.~(\ref{eq:photon-rate}) is plagued by
  very serious pathologies that appear as ultraviolet divergences.
  Firstly, the r.h.s. of eq.~(\ref{eq:photon-rate}) turns out to be
  infinite at any fixed photon energy $\omega$ due to some unphysical
  vacuum contributions, i.e. processes where a photon is produced
  without any particle in the initial state. Secondly, the remaining
  terms, even if they give a finite photon production rate, lead to an
  energy dependence of this rate which is too hard for being
  integrable: one would conclude based on this formula that the total
  energy radiated as photons per unit time by a finite volume is
  infinite, which clearly violates energy conservation.
  
  It was claimed in \cite{BoyanV1} that the vacuum terms could be
  discarded simply by subtracting to the r.h.s. of
  eq.~(\ref{eq:photon-rate}) the same formula evaluated in the vacuum.
  This indeed has the desired effect, but is a totally {\it ad hoc}
  prescription because nowhere in the derivation of the formula
  appears this subtraction term. Or said differently, since
  eq.~(\ref{eq:photon-rate}) is a direct consequence of the definition
  of eq.~(\ref{eq:N1}), whatever is wrong with the final formula
  signals a problem either with this definition or with the model.
  
  Similarly, the authors of \cite{BoyanV1} suggested that the
  divergence that appears in the total radiated energy can be
  subtracted by multiplying the creation and annihilation operators
  used in the definition of the number of photons by some wave
  function renormalization constants. However, no such constants
  appear in the derivation: the operators $a_{\rm out}, a^\dagger_{\rm
    out}$ can be related to their ``in'' counterparts directly by
  means of the $S$-matrix.
\end{itemize}

\section{Conclusions}
In this paper, we have obtained a new simple derivation of a formula
recently proposed in order to calculate the number of photons produced
by a system during a finite time. The purpose of this derivation is to
be sufficiently transparent in order to exhibit all the underlying
hypothesis.  In particular, in order to obtain this formula, one would
have to impose that the electromagnetic interactions be turned off
before the initial time $t_i$ and after some final time $t_f$. This is
clearly unphysical and not surprisingly leads to serious pathologies.
Our conclusion is that eq.~(\ref{eq:photon-rate}) does not make sense
when keeping a finite $t_f-t_i$ time interval. The transient effects
associated to a finite lifetime of the hot and dense system, which
could enhance the photon production as suggested in \cite{BoyanV1},
cannot be evaluated that way.

\section*{Acknowledgments}
We would like to thank D.Boyanovsky, I.~Dadic, G.D.~Moore and
J.~Serreau for discussions. E.S.F. would like to thank the members of
Laboratoire de Physique Th\'eorique, where this work was initiated,
for their kind hospitality.  E.S.F. is partially supported by CAPES,
CNPq, FAPERJ and FUJB/UFRJ.

\bibliographystyle{unsrt}
%\bibliography{biblio}

\end{document}